\documentclass[twocolumn,showpacs,preprintnumbers,amsmath,amssymb,a4paper,superscriptaddress]{revtex4}
\usepackage{amsmath}
\usepackage{graphicx}
\usepackage{dcolumn}
\usepackage{bm}

\begin{document}

\title{Shifts and widths of Feshbach Resonances in Atomic Waveguides}
\date{\today}
\author{Shahpoor Saeidian}
\email[]{saeidian@iasbs.ac.ir}
\affiliation{Optics and Photonics Research Center, Department of Physics, Institute for Advanced Studies in Basic Sciences
(IASBS),
Gava Zang,
Zanjan 45137-66731,
Iran
}%
\author{Vladimir S. Melezhik}
\email[]{melezhik@theor.jinr.ru} \affiliation{Bogoliubov Laboratory of Theoretical Physics, Joint Institute for Nuclear Research, Dubna, Moscow Region 141980, Russian Federation}%
\author{Peter Schmelcher}
\email[]{peter.schmelcher@physnet.uni-hamburg.de}
\affiliation{Zentrum f\"ur Optische Quantentechnologien, Universit\"at Hamburg, Luruper Chaussee 149, 22761 Hamburg, Germany}%

\date{\today}
\begin{abstract}\label{txt:abstract}
We develop and analyze a theoretical model which yields the shifts and widths of Feshbach
resonances in an atomic waveguide. It is based on a multichannel approach for
confinement-induced resonances (CIRs) and atomic transitions in
the waveguides in the multimode regime. We replace in this scheme the
single-channel scalar interatomic interaction by the four-channel tensorial potential modeling resonances of
broad, narrow and overlapping character according to
the two-channel parametrization of A.D.Lange et. al. \cite{Lange}.
As an input the experimentally known parameters of Feshbach
resonances in the absence of the waveguide are used. We
calculate the shifts and widths of s-, d- and g-wave magnetic
Feshbach resonances of Cs atoms emerging in harmonic
waveguides as CIRs and resonant enhancement of the transmission at zeros of the free space scattering length.
We have found the linear dependence of the width of the resonance on the
longitudinal atomic momentum and quadratic dependence on the waiveguide
width. Our model opens novel possibilities for quantitative studies of the scattering
processes in ultracold atomic gases in waveguides beyond the framework of s-wave resonant scattering.
\end{abstract}

\maketitle

\section{INTRODUCTION}
Impressive progress of the physics of ultracold quantum gases have opened new pathways for the
study of low-dimensional few-body systems (see, for example \cite{Chin2010,Kohler2006})
as well as strongly correlated many-body systems \cite{Bloch2008,Yurovsky2008}).
Specifically, it was shown that the confining geometry of atomic traps can drastically change
the scattering properties of ultracold atoms and induce resonances in the collisions (confinement-induced resonances (CIRs)) \cite{Olshanii1}.
The CIR for bosons has been found to occur in the vanishing energy collisional limit
as a consequence of the coincidence of the binding energy of a diatomic molecular state
with the energy spacing between the levels of the confining (harmonic)
potential \cite{Olshanii2, Melezhik07,Saeidian}. It was shown that this coincidence leads
to a divergence of the effective interatomic coupling constant $g_{1D}$ and to a total atom-atom reflection, which appears as
a broad dip in the transmission coefficient T, thereby creating a gas of impenetrable bosons \cite{Girardeau}.   CIRs have also been extensively
studied e.g. in the context of three-body \cite{Mora1, Mora2}, and four-body \cite{Mora3} scattering in confining traps, fermionic p-wave
scattering \cite{Granger}, distinguishable atom scattering \cite{Kim2, Kim3, Melezhik07, Melezhik09} or multichannel scattering \cite{Saeidian, Melezhik11,Olshanii3} in atomic waiveguides.
Two novel effects were predicted for distinguishable atoms: the so-called dual CIR yielding
a complete suppression of quantum scattering \cite{Kim2}, and the resonant molecule formation in tight waveguides \cite{Melezhik09}.
D-wave resonant scattering of bosons in confining harmonic waveguides has been analyzed very recently in \cite{Giannakeas}.
Remarkable experimental progress has lead to the observation of CIRs,
for identical bosons \cite{Kinoshita, Paredes,Haller2009,Haller2010} and fermions\cite{Guenter,Moritz,Kohl2011}, as well as to
distinguishable atoms \cite{Lamporesi2010}.

However, the recent experimental \cite{Lamporesi2010,Haller2010,Kohl2011,Haller2011}
and theoretical \cite{Peng2010,Zhang2,Melezhik11,Sala2011} investigations of the CIRs clearly
show that, despite the impressive progress, the existing theoretical models of CIRs need to be improved for
a quantitative description of the experiments in this field.
The obvious drawback of the existing theoretical models for CIRs in atomic waveguides is the single-channel character of the
simple interatomic interactions employed. Single-channel potentials with zero-energy
bound states were used for simulating magnetic Feshbach resonances in free space representing a
necessary ingredient for the appearance of the CIR in a confining trap(see, for example \cite{Chin2010,Yurovsky2008}). In the seminal work of
Olshanii \cite{Olshanii1} and in subsequent papers \cite{Olshanii2,LowDSystems}, including the recent works
\cite{Peng2010,Zhang2}, the simple form of a pseudo-potential was used modeling the interatomic interactions. More realistic potentials
were used in our previous works \cite{Kim2,Kim3,Saeidian,Melezhik09,Giannakeas,Melezhik11} as well as
in the works of other authors \cite{Olshanii2,Granger,Sala2011,Grishkevich} which, however, all possess a single-channel character.
The single-channel interatomic interaction approach permits to explore only the main attribute of the
Feshbach resonances in the 3D free space, namely the appearance of a singularity in the s-wave
scattering length $a_s \rightarrow\pm\infty$ when the molecular bound state with energy $E_B$ crosses the atom-atom
scattering threshold at energy $E=0$ in the entrance channel. However, other important
parameters of the Feshbach resonance, such as the rotational and spin structure of
the molecular bound state in the closed channel as well as the width
$\Delta$ of the resonance characterizing the coupling $\Gamma$ of the molecular state with the
entrance channel, were ignored. The main goal of the present work is to extend these theoretical approaches developed
earlier for the CIRs and transverse excitations/deexcitation processes for collisions
in harmonic waveguides \cite{Saeidian} to the case of a tensorial interaction
as well as taking into account the width of the magnetic Feshbach resonances responsible for the CIR. The parameters
obtained from the experimental analysis of the magnetic Feshbach resonances in free space,
namely the resonant energies $E_{c,i}$ (or the corresponding values of the field strengths $B_{c,i}$ of the external magnetic field),
the widths of the resonances $\Delta_i (\Gamma_i)$, spin characteristics and the background scattering length $a_{bg}$,
are used as input parameters of our model.

It is well-known that the most adequate computational schemes
including all the above mentioned parameters of the Feshbach
resonances are multi-channel scattering
\cite{Chin2010,Chin2004,Lange} and multi-channel quantum-defect
\cite{Mies2000} approaches providing a quantitative description
for a broad regime of experimental parameters for the Feshbach
resonances in ultracold atomic gases. In the present work the
two-channel potential scheme \cite{Lange}, permitting the
efficient modeling of the interatomic interaction near the known
magnetic Feshbach resonances of the Cs gas in free space
\cite{Chin2004}, is included in our multichannel approach
\cite{Saeidian} instead of the simplified interparticle
single-channel potential we have used so far for analyzing CIRs
and transverse excitations in atomic waveguides. With this
approach we model s-, d- and g-wave Feshbach resonances in
ultracold Cs (s-,d- and g- indicate here the "exit" channels, i.e.
they correspond to rotational quantum numbers of the molecular
states in the closed channels), which were observed in free space
scattering experiments \cite{Chin2004}, and quantitatively analyze
the shifts and widths of the resonances in the harmonic waveguides
at experimental conditions closed to the ones encountered in the
works \cite{Haller2009,Haller2010}. We note that selected aspects
of the modeling of magnetic Feshbach resonances in  optical traps and lattices
were also considered in \cite{Dickerscheid,Diener,Yurovsky2006,Schneider,Peng}.

\section{FESHBACH RESONANCE MODEL}

We consider the collision of two identical bosonic atoms in a harmonic waveguide.  This two-body problem permits the separation of the center-of-mass
and relative motion yielding the following Hamiltonian for the relative atomic motion
\begin{equation}
\hat{H}(r,\theta)=[-\frac{\hbar ^2}{2\mu}\nabla^2+ \frac{1}{2}\mu\omega_{\perp}^2\rho^2]\hat{I} + \hat{V}(r)
\end{equation}
with $\rho = r\sin\theta$ and the trap potential $1/2\mu\omega_{\perp}^2\rho^2$.
$\hat{V}(r)$ is the 4-channel interatomic potential and $\hat{I}$ is the unit matrix,
$r$ is the relative radial coordinate and $\mu = m/2$ is the reduced mass of the atoms.

Following the scheme suggested in \cite{Lange} for describing the three magnetic Feshbach resonances in an ultracold Cs gas let us suppose
that initially  the scattering atoms are prepared in one spin configuration $|e\rangle$ (called the
``entrance channel") and the ``closed channels" $|c,i\rangle$ (i=1, 2, 3) support s-, d- and g-wave molecular bound states.
at -11.1G, 47.78G and 53.449G.
The quantum state of an atomic pair with energy $E$ is described as
$$|\psi\rangle=\sum_{i=1}^3\psi_{c,i}({\bf r})|c,i\rangle+\psi_{e}({\bf r})|e\rangle$$
satisfying the Schr\"{o}dinger's equation with the Hamiltonian (1).  A four-channel square-well potential
\begin{equation}
\hat{V}=\left (\begin{array}{cccc} -V_{c,3}& 0 & 0 & \hbar\Omega_3\\ 0 & -V_{c,2} & 0 &  \hbar\Omega_2\\
  0&0&-V_{c,1}  & \hbar\Omega_1\\ \hbar\Omega_3 &  \hbar\Omega_2 & \hbar\Omega_1& -V_e
\end{array}\right )  \begin{array}{cc} (if & r<\overline{a})\end{array}\\
\end{equation}
\begin{equation}\nonumber
=\left (\begin{array}{cccc} \infty& 0 & 0 &0\\ 0 & \infty & 0 & 0\\
  0&0& \infty&0\\0 & 0&0& 0
\end{array}\right )  \begin{array}{cc} (if & r>\overline{a})\end{array}
\end{equation}
is employed to describe the colliding atoms in the ``entrance channel"
$|e\rangle$ and the weakly-bound molecules in the ``closed channels" $|c,i\rangle$ near a Feshbach
resonance. For $r< \overline{a}=4\pi\Gamma(1/4)^{-2}R_{vdW}$, we assume the attractive potential can support multiple molecular states - that is,
$ V_{e}$,$ V_{c_{i}}\gg E_{vdW}=\hbar^2/2mR_{vdW}^2$ - and $\hbar\Omega_i$ induce Feshbach couplings
between the channels \cite{Lange}. The size $\overline{a}$ of the potential action is chosen to account for the interatomic interaction
determined by the van-der-Waals (vdW) tail with the length $R_{vdW}=1/2(2mC_6/\hbar^2)^{1/4}$ and the energy
scale $E_{vdW}=\hbar^2/2mR_{vdW}^2$ \cite{Gribakin},
where $C_6$ is the corresponding van der Waals coefficient and $\Gamma(x)$ is the gamma function. For $r> \overline{a}$, entrance- and closed-channel
thresholds are set to be $ E=0$ and $E=\infty$, respectively (See Fig.1 in Ref.\cite{Lange}).

Such a choice of the interatomic interaction permits a simple
parametrization of the atom-atom scattering in universal terms
of the energy of the i-th bare bound state $E_i$,  the Feshbach
coupling strength of the bound molecular state with the entrance
channel $\Gamma_i$ and the background scattering length $a_{bg}$,
which is convenient for an analysis of experimental data near
magnetic Feshbach resonances \cite{Chin2010,Lange}. When the
mixing between the closed channels and the entrance channel is
weak and the background scattering length $ |a_{bg}|$ considerably
exceeds the range of the interatomic interaction $\overline{a}$, the
s-wave scattering length $a$ and the binding energy $E_b$ in free space are given by \cite{Lange}:
\begin{equation}
\frac{1}{a-\overline{a}}=\frac{1}{a_{bg}-\overline{a}}+\frac{1}{\overline{a}}\sum_{i=1}^{3}\frac{\Gamma_i/2}{E_i},
\end{equation}
and
\begin{equation}\label{binding energy}
E_b=\frac{\hbar^2k_m^2}{2\mu}\,, \,\,\, k_m=\frac{1}{a_{bg}-\overline{a}}+\frac{1}{\overline{a}}\sum_{i=1}^{3}\frac{\Gamma_i/2}{E_b+E_i}
\end{equation}
respectively.
Then, assuming that the bare bound  states can be linearly tuned magnetically by a linear Zeeman shift -namely, $ E_{i}=\delta\mu_{i}(B-B_{c,i})$, where $\delta\mu_{i}$
is the relative magnetic moment between the entrance and i-th channels and $ B_{c,i}$ is the crossing field value of the $i$th bare bound
state, the scattering length $a(B)$ can be represented as
\begin{equation}\label{Scatt length eq}
\frac{a}{a_{bg}}=\prod^{3}_{i=1}\frac{B-B^*_i}{B-B_{0,i}}\,\,.
\end{equation}
Here $B_{0,i}$ is the $i$th lowest pole of $a$ and $B^*_{i}$ the $i$th lowest zero.  The width of the $i$th Feshbach resonance can be defined as
$\Delta_{i}=B_{i}^*-B_{0,i}$.  The binding energy of Feshbach molecules can be measured in the
experiment by e.g. radio frequency and microwave
spectroscopy \cite{Regal, Bartenstein, Claussen, Thompson, Papp}.  Using the fitting parameters $a_{bg}$, $\delta\mu_{i}$, $\Gamma_{i}$ and $B_{c,i}$,
one can fit the experimental data \cite{Regal, Bartenstein, Claussen, Thompson, Papp} with  (\ref{binding energy}), from which one can calculate
$a(B)$ (3), $B^*_i$ and $B_{0,i}$ (5) \cite{Lange}.

\begin{figure}[!t]
\includegraphics[width=1.1\columnwidth]{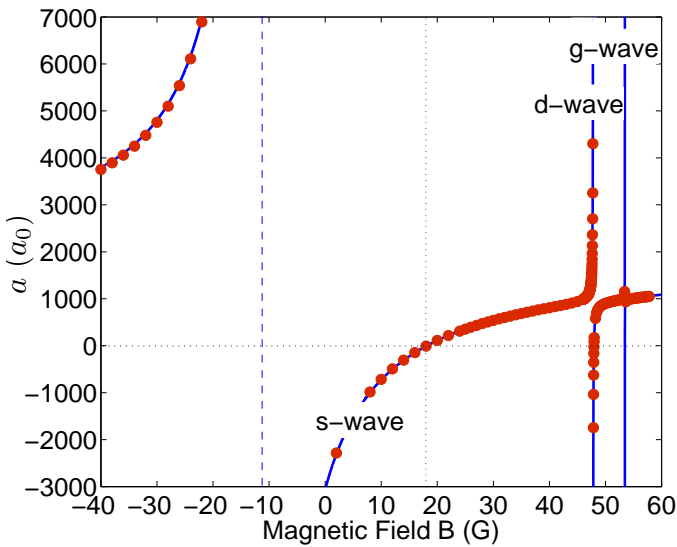}
\caption{(Color online) The s-wave scattering length $a$ of  $|F=3, m_F=3\rangle$ Cesium atoms as a function of the magnetic field  $B$.  The solid curve shows the analytical result (5) and the
dots show the numerical result (see the text).}
\label{fig1}
\end{figure}

\begin{figure}[hbt]
\includegraphics[width=1.1\columnwidth]{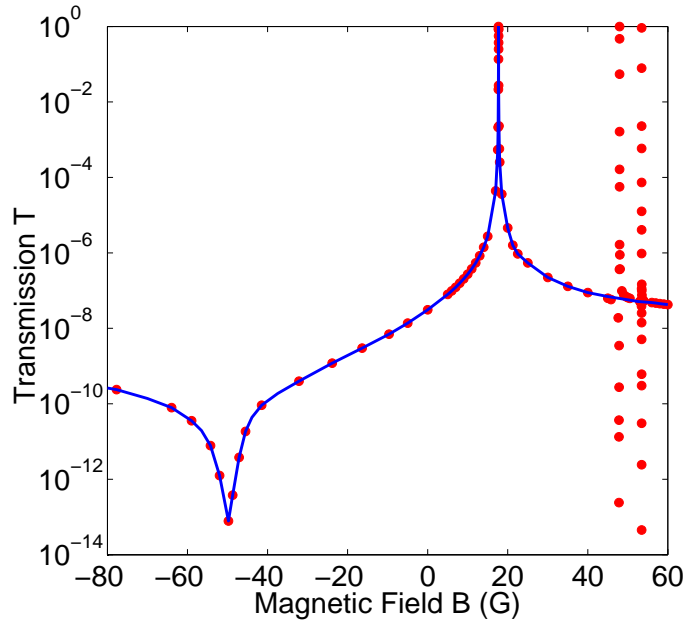}
\caption{(Color online) The transmission coefficient $T$ for the harmonic trap with $\omega_{\perp}=14.9$ kHz as a function of the
magnetic field $B$ for the case of the 2-channel (solid curve) and
4-channel (dots) potential (2).} \label{fig2}
\end{figure}

The result of the procedure is given in the  Table I of ref.
\cite{Lange} for ultracold $Cs$ Feshbach molecules where  $
a_{bg}=1875 a_{0}$ and $ \overline{a} = 95.7 a_{0}$ ($a_0$ is the
Bohr radius) are also given. With these values for $B_{0,i}$ and $B_{i}^*$ we
show the scattering length $a(B)$ as a function of the
magnetic field $B$ according to Eq.(\ref{Scatt length eq}) in Fig.1.

The parameters $ a_{bg}$, $\delta\mu_{i}$, $\Gamma_{i}$, $B_{c,i}$, $B^*_i$ and $B_{0,i}$ from
\cite{Lange} together with $a(B)$ defined by Eq.(5) are used for fitting
the diagonal terms $V_{c,i}$ and $V_e$ in the tensor potential (2). The nondiagonal terms $\hbar\Omega_i$
are defined by the formulas \cite{Lange}
$$\Gamma_{i}/2=2\theta^{2}V_{c,i}\,,\,\,\,
\tan 2\theta_i=\frac{2\hbar\Omega_i}{V_e-V_{c,i}}\,\,.$$ The
scattering length $a(B)$ is then calculated for different $B$ and varying parameters of the potential $\hat{V}$ by solving the
Schr\"odinger equation
\begin{equation}
\left ([-\frac{\hbar^2}{2\mu}\nabla^2+ \frac{1}{2}\mu\omega_{\perp}^2\rho^2]\hat{I}+\hat{B} +\hat{V}(r)\right )|\psi\rangle=E|\psi\rangle
\end{equation}
in free space ($\omega_{\perp}=0$) with the scattering boundary conditions
$$\psi_e({\bf r})\rightarrow \exp\{ikz\} + f(k,\theta)/r\exp\{ikr\}\,,\,\,\, \psi_{c,i}({\bf r})\rightarrow 0$$
at $kr\rightarrow\infty$ for the fixed $E\rightarrow 0$  $ (k=\sqrt{2\mu E}/\hbar\rightarrow 0)$ \cite{MelHu}. The diagonal
matrix $\hat{B}$ in (6) is defined as $B_{ii}=\delta\mu_i(B-B_i)$ $(i=1,2,3)$ and $B_{ee}=0$.
After separation of the angular part in (6) we come to the system of four coupled radial equations
\begin{equation}
[-\frac{\hbar^2}{2\mu}\frac{d^2}{dr^2} -\frac{\hbar^2l_{\alpha}(l_{\alpha}+1)}{2\mu r^2} +B_{\alpha\alpha}]\phi_{\alpha}(r) + \sum_{\beta}V_{\alpha\beta}(r)\phi_{\beta}(r)= E\phi_{\alpha}(r)
\end{equation}
for the radial part $\phi_{\alpha}(r)$ of the desired wave
function
$|\psi\rangle=\sum_{\alpha}\psi_{\alpha}({\bf{r}})|\alpha\rangle =
\sum_{\alpha}\phi_{\alpha}(r)Y_{l_{\alpha}0}(\hat{r})|\alpha\rangle$,
where $\alpha=\{e,i=1,2,3\}$, $(l_e=0, l_1=0, l_2=2, l_3=4)$ and the matrix
elements $V_{\alpha\beta}(r)$ are defined by Eq.(2). The
centrifugal barrier  $-\frac{\hbar^2l_{\alpha}(l_{\alpha}+1)}{2\mu
r^2}$ in (7) models at $r\rightarrow 0$ the correct asymptotic
behavior $\phi_{i}(r) \sim r^{l_i+1}$ of the molecular bound
states $|c,i\rangle$ in the closed channels which couple to the
entrance s-wave ($l_e=0$) channel $|e\rangle$ by the nondiagonal terms
$V_{\alpha\beta}(r) (\alpha\not=\beta)$.

By varying the
$V_{c,i}$, $V_e$ and $\Omega_i$ we obtain an excellent agreement of the
calculated s-wave scattering length $a(B)$ with the analytical results from
\cite{Lange} for Cesium atoms in the hyperfine state $|F=3, m_F=3\rangle$ for the considered magnetic
field regime $-40G<B<60G$ (see dots in Fig.1).  In this regime,
we observe three resonance terms, which correspond to the coupling to the s-, d-, and g-wave molecular states.

Next we analyze the scattering properties of the s-, d- and g-wave
magnetic Feshbach resonances in harmonic waveguides by integrating the Schr\"odinger equation (6)
for $\omega_{\perp}\neq 0$ with the scattering boundary conditions
\begin{equation}
\psi_e({\bf r}) = \left (\cos (k_0 z) + f_e \exp\{ik_0 \mid z\mid\}\right
)\Phi_{0}(\rho)\,\,,\,\,\, \psi_{c,i}({\bf r})\rightarrow 0
\end{equation}
at $\mid z\mid = \mid r\cos\theta\mid \rightarrow \infty$ adopted
for a confining trap \cite{Saeidian}. $f_e(E)$ is the
scattering amplitude, corresponding to the symmetry with respect to the exchange $z
\rightarrow -z$ (we consider collisions of identical bosonic Cs
atoms), $\Phi_0(\rho)$ is the wave function of the ground-state of
the  two-dimensional harmonic oscillator  and $k_0=\sqrt{2\mu
(E-\hbar\omega_{\perp})}/\hbar=\sqrt{2\mu E_{\parallel}}/\hbar$.
In the presence of of the harmonic trap ($\omega_{\perp}\not=0$) the problem (6),(8) becomes non-separable
in the plane $\{\rho,z\}$, i.e. the azimuthal angular part is
separated in the wave function $|\psi\rangle$ and Eq.(6) is
reduced to the coupled system of four 2D Schr\"{o}dinger-type
equations. To integrate  this coupled
channel 2D scattering problem in the plane $\{r,\theta\}$ we have extended the computational
scheme \cite{Saeidian}.

The computations have been performed in a range of variation of
$\omega_{\perp}$ close to the experimental values of the trap
frequencies $\sim 2\pi \times 14.5$kHz \cite{Haller2010}. We have
integrated Eq.(6) for varying $B$ and fixed longitudinal colliding
energy $E_{||}=E-\hbar\omega_{\perp}$. In the main part of
computations the energy $E_{||}$ was chosen very low
$E_{||}=1.0\times 10^{-16}(\frac{\hbar^2}{\mu\overline{a}^2})$
$\rightarrow 0$ $(k_0=1.0\times 10^{-8}(\frac{1}{a_0})\rightarrow
0)$ to have a possibility for direct comparison with existing
pseudopotential estimates obtained in zero-energy limit. The
integration was performed in the units of the problem leading to
the scale transformation: $r \rightarrow \frac{r}{\overline{a}}$,
$E \rightarrow \frac{E}{E_0}$, $V \rightarrow \frac{V}{E_0}$, and
$\omega_{\perp} \rightarrow \frac{\omega_{\perp}}{\omega_0}$ with
$E_0=\frac{\hbar^2}{\mu\overline{a}^2}$, and
$\omega_0=\frac{E_0}{\hbar}$.

\section{RESULTS AND DISCUSSION}

\subsection{Transmission coefficient}
In Fig.2 we present the transmission coefficient $T(B)=|1+
f_e(B)|^2$ as a function of B calculated for the harmonic trap
$\omega_{\perp}=14.9$kHz with the two-channel (i=1) and
four-channel (i=1,2,3) character of the tensorial
interaction $\hat{V}(r)$ (2).
\begin{figure}[t]
\includegraphics[width=1.\columnwidth]{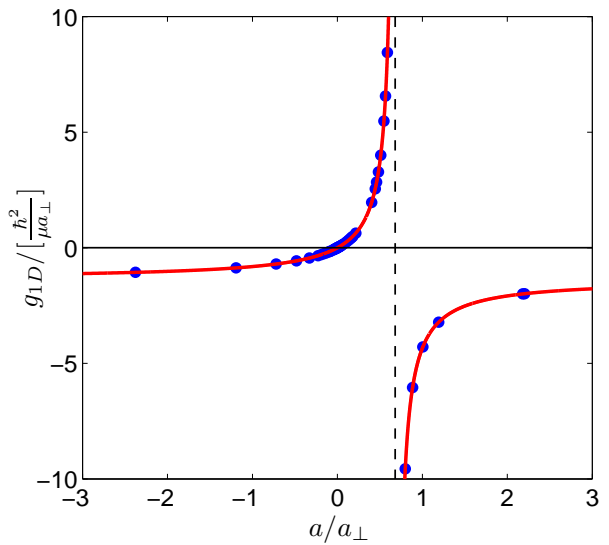}
 \caption{(Color online) The coupling constant $g_{1D}$ as a function
of the scattering length $a$ in free space in the region
near the s-wave Feshbach resonance, calculated for the harmonic trap with $\omega_{\perp}=14.9$kHz. Dots indicate the numerical
data and the solid curve the analytical results with
the formulas derived in the pseudopotential approach
\cite{Olshanii1}.} \label{fig3}
\end{figure}
The two-channel potential $\hat{V}(i=1)$ supports only one broad
s-wave Feshbach resonance
while the four-channel potential
$\hat{V}(i=1,2,3)$ supports all three s-, d- and g-wave magnetic
Feshbach resonances in free space ($\omega_{\perp}=0$) at
the fields $B_{0,1}=-11.1$G, $B_{0,2}=47.78$G and $B_{0,3}=53.449$G,
respectively \cite{Lange}. All these resonances are also developed
in the calculated curve $T(B)$ of the transmission coefficient in
the trap ($\omega_{\perp}\neq 0$).

\begin{figure}[hbt]
\includegraphics[width=1.\columnwidth]{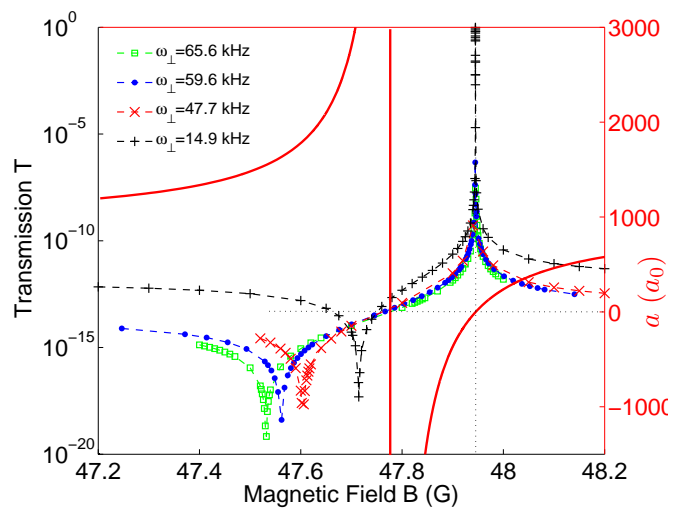}
\caption{(Color online) The transmission coefficient $T(B)$ as a function of the
strength of the magnetic field for several trap frequencies
$\omega_{\perp}$. Here the s-wave scattering length $a(B)$ from
Fig.1 is also given (solid line).} \label{fig4}
\end{figure}
\begin{figure}[hbt]
\includegraphics[width=1.\columnwidth]{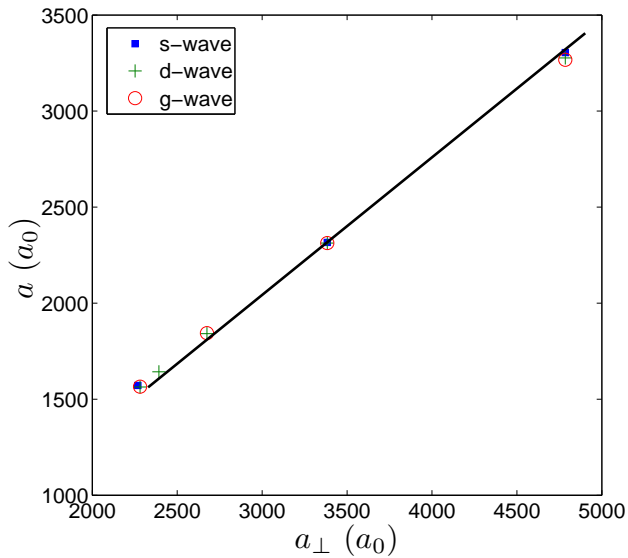}
\caption{(Color online) The dependence of the s-wave scattering length $a(B)$ on
the width $a_{\perp}$ of the waveguide at the points $B_{min}$ of the minimum of the
transmission coefficient $T(a(B),a_{\perp})$ (see Fig.4).
The dots, pluses and cycles correspond to the calculated points
near the s-,d- and g-wave magnetic Feshbach resonances, respectively. The solid
curve corresponds to the formula $a=a_{\perp}/C$
\cite{Olshanii1}.} \label{fig5}
\end{figure}

\begin{figure}[hbt]
\includegraphics[width=1.1\columnwidth]{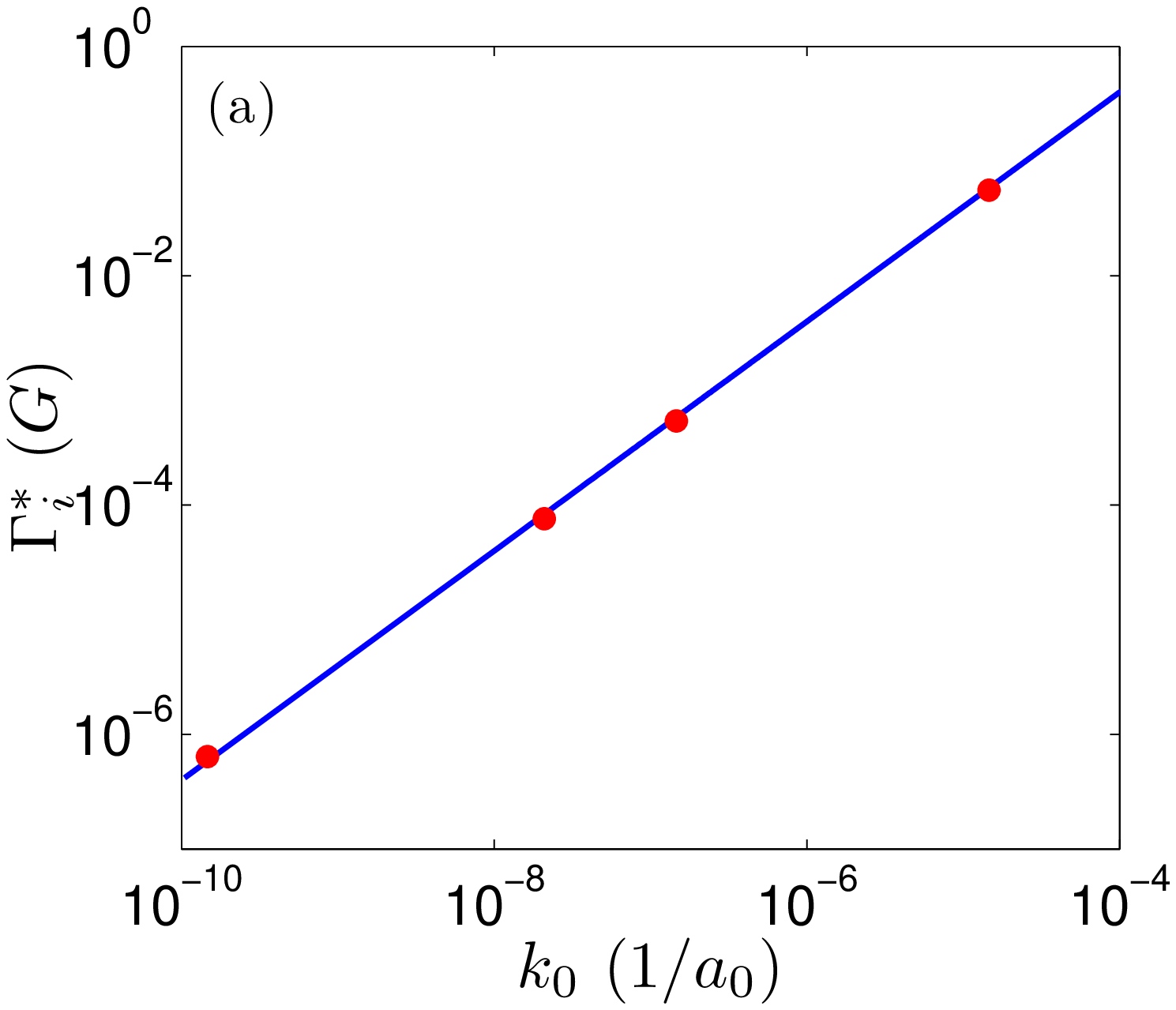}
\includegraphics[width=1.\columnwidth]{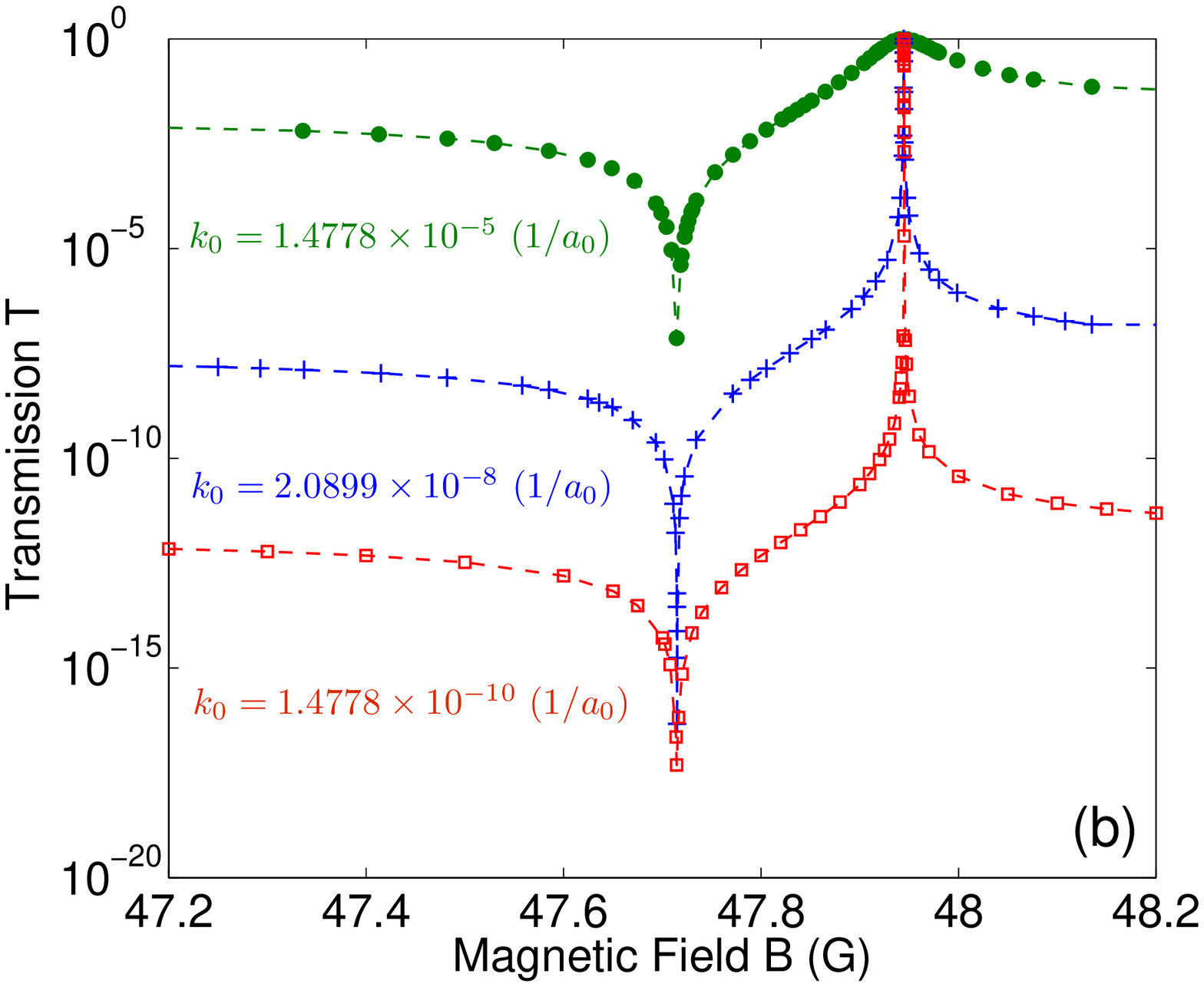}
\caption{(Color online) (a) The dependence of the width $\Gamma_i^*$ on the
longitudinal momentum  $k_0$ near the d-wave Feshbach resonance.
The solid line has been obtained via Eq.(10), solid circles
indicate the widths $\Gamma_i^*(k_0)$ extracted from the
numerically calculated $T(B,k_0)$.(b) The transmission coefficient
T as a function of the magnetic field B calculated for a few
$k_0$. For both subfigures $\omega_{\perp}$=14.9kHz.}
\label{fig55a}
\end{figure}

\begin{figure}[hbt]
\includegraphics[width=1.\columnwidth]{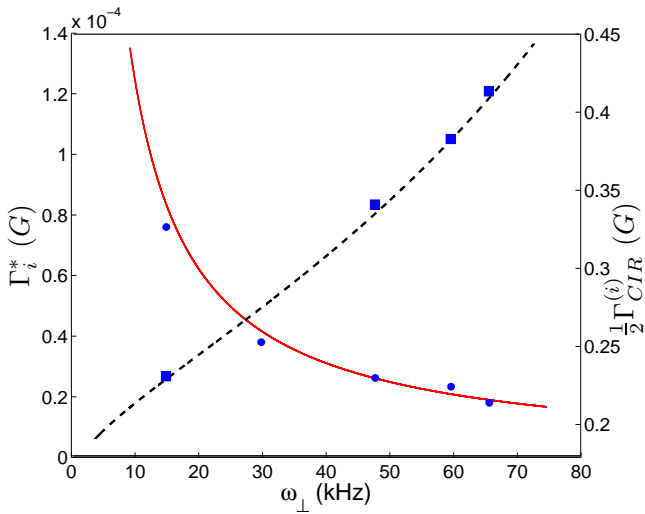}
\caption{(Color online) The dependence of the width $\Gamma_i^*$ on the trap
frequency $\omega_{\perp}$ near the d-wave Feshbach resonance at
$k_0$=2.0899$\times 10^{-8}(\frac{1}{k_0})$. The solid curve has been obtained via Eq.(10), solid circles indicate the widths
$\Gamma_i^*(\omega_{\perp})$ extracted from the numerically calculated
$T(B,\omega_{\perp})$.   The analytical (dashed line) and numerical (squares) results for the right half width $1/2\Gamma^{(i)}_{CIR}$
versus $\omega_{\perp}$ are also given.} \label{fig55b}
\end{figure}

First, we have analyzed the region of $B\simeq-11.1$ G near the
s-wave resonance. In Fig.3 we show the calculated 1D coupling
constant $g_{1D}=\lim_{k_0\rightarrow
0}Re\{f_e(k_0)\}/Im\{f_e(k_0)\}k_0/\mu$ \cite{Olshanii2} as a
function of the scattering length $a(B)$ in this region.  By
tuning $B$ in the interval $-60G \leq B\leq 40G$ , the s-wave
scattering length $a(B)$ changes from $-\infty$ to $+\infty$. At
$B=-49.78$G corresponding to the point $a(B)=a_{\perp}/C $ (where
$C=1.4603..$ and $a_{\perp}=\sqrt{\hbar/\mu\omega_{\perp}}$) of
the appearance of the CIR \cite{Olshanii1}, the coupling constant
$g_{1D}(a/a_{\perp})$ diverges and the behavior is in very good
agreement with our previous computation of $g_{1D}(a/a_{\perp})$
performed with the single channel screened Coulomb potential
\cite{Saeidian} and the formulas $g_{1D}=2\hbar a/(\mu
a_{\perp}^2)/(1-C a/a_{\perp})$ derived in a pseudopotential
approach \cite{Olshanii1}.

Let us next analyze the region of the d-wave magnetic Feshbach
resonance near the point $47.78$G (see Fig.1) - the region of
major experimental interest due to atomic loss and the
formation of Cs molecules in the confined ultracold gas of Cs
atoms \cite{Haller2009,Haller2010}. The results of our
computations are illustrated by the curves $T(B)$ calculated for
different $\omega_{\perp}$ (Fig.4) corresponding to the transverse
frequencies of the optical trap being used in the
experiment \cite{Haller2010}. It is clearly shown that the
position $B_{min}$ of the transmission coefficient $T(B)$ minimum
is dependent on the trap width
$a_{\perp}=\sqrt{\hbar/(\mu\omega_{\perp})}$ and the corresponding
scattering length $a(B_{min})$ at the point $B_{min}$ is
accurately described by the formulas $a(B_{min}) = a_{\perp}/C$
obtained by Olshanii \cite{Olshanii1} for the position of the CIR
in a harmonic waveguide. The latter fact is illustrated in Fig.5. Here
the results of the calculation of the dependence of $a(B_{min})$
on  $a_{\perp}$ for s-wave and g-wave Feshbach resonances are also
given. This analysis clearly demonstrates that the law
$a=a_{\perp}/C$ for the position of the CIR in a harmonic
waveguide is fulfilled with high accuracy for the Feshbach
resonances of different tensorial structure, although the law was
initially obtained at zero-energy limit in the framework of an s-wave single-channel
pseudopotential approach to the interatomic interaction
\cite{Olshanii1}. We also see that the positions $B_{max}$ of the
maximums of the coefficients $T(B)$ calculated for different
$\omega_{\perp}$ are independent
of $\omega_{\perp}$ and coincide at the point $B_{2}^*=47.944$G of the zero of the scattering length in
free space ($a(B_{2}^*)=0$). This fact is in agreement with
the analytic result $T=\mid 1+f_e\mid^2\rightarrow 1$ obtained in
the pseudopotential approach at $a\rightarrow 0$. The same
behavior of the $T(B)$ coefficients has been found in the
vicinity of 18.1G and 53.46G (points $B_1^*$ and $B_3^*$ of s-wave
scattering length zeros near the s- and g-wave Feshbach
resonances).

\subsection{Width of the resonant enhancement of the transmission}

While the coefficients $T(B)$ in the vicinity of the points $B_i^*$ (positions of the zeros of the scattering length $a(B)$)
show a resonant behavior one can define the width $\Gamma_{i}^*$ of this resonance as the width at half $T$ maximum.
By using the formulas for the even scattering amplitude in the trap \cite{Schmiedmayer,Olshanii3}
$$
f_e=-\frac{1}{1+i\cot\delta_{1D}} = -\frac{1}{1+i k_0 a_{1D}}
$$
valid near the CIR in the zero-energy limit and the definitions $a_{1D}=\frac{a_{\perp}}{2}(C-\frac{a_{\perp}}{a})$ and
$a=a_{bg}\prod^{3}_{i=1}\frac{(B-B_i^*)}{(B-B_{0,i})}$,
we obtain
\begin{equation}
\Gamma_{i}^* = \frac{4 a_{bg}\gamma_i a_{\perp}^2k_0\Delta_i}{4
a_{bg}^2 \gamma_i^2-a_{\perp}^4k_0^2(1- C\gamma_i
\frac{a_{bg}}{a_{\perp}})^2} \,\,,
\end{equation}
where $\gamma_i=\prod_{j\neq
i}^3\frac{(B_i^*-B_j^*)}{(B_i^*-B_{0,j})}$. The above formula is
valid for the condition $\Gamma_i^*\ll B_i^*$ of narrow resonances,
which is fulfilled with high accuracy for the d- and g-wave resonances
and less accurately for s-wave resonance (see Fig.2). In the low
energy limit $k_0\rightarrow 0$ the expression is reduced to
\begin{equation}
 \Gamma_{i}^* = \Delta_i \frac{a_{\perp}^2 k_0}{a_{bg}\gamma_i} =
 \Delta_i\frac{\sqrt{2E_{\parallel}}}{\sqrt{\mu}a_{bg}\omega_{\perp}\gamma_i} \,\,\,,
\end{equation}
where $\Delta_i = B_i^*-B_{0,i}$ is the width of the Feshbach
resonance in free space and the dimensionless  $\gamma_i$- factor
$\approx\frac{1}{2}$ for the case of d- and g-wave resonances and
$\approx 1$ for s-wave resonance. Figs.6 and 7 demonstrate the
very good agreement of these formulas with the numerical
computations. Actually, in Fig.6(a) one can see in a broad range
of $k_0$ variation the very good agreement of the width $\Gamma_i^*(k_0)$,
extracted from numerically calculated $T(B,k_0)$, with the linear
functional dependence on the longitudinal momentum $k_0$, following from
Eq.(10). Also, the $\Gamma_i^*(\omega_{\perp})$ has inverse dependence on
$\omega_{\perp}$. It is confirmed by results of numerical calculations of
$\Gamma_i^*(\omega_{\perp})$ given in Fig.7. Thus, Eq.(10) can be
used for extracting important information about the analyzed
system. Indeed, by measuring the width $ \Gamma_{i}^*$ one can
extract from Eq.(10) the longitudinal momentum $k_0=\sqrt{2\mu
E_{\parallel}}/\hbar$ (longitudinal colliding energy
$E_{\parallel}$) and estimate the ``longitudinal" temperature of
the atomic cloud in the trap. This expression also shows that one
can control the width $\Gamma_{i}^*$ of the resonance by varying
the trap frequency $\omega_{\perp}$. Increasing of
$\omega_{\perp}$ leads to a narrowing of the resonance (see Figs.4
and 7), the effect of which can be used experimentally. Fig.6(b)
demonstrates the stability of the position of the CIR (the minimum
of the transmission coefficient $T$) with respect to $k_0$ variation and the linear growing of the
transmission coefficient $T$ with $k_0$ increasing. It is shown
that already at very low $k_0\sim 10^{-5}(\frac{1}{a_0})$ the
$T$-coefficient becomes large enough and consequently can be
experimentally ``visible" near CIR.
\begin{figure}[hbt]
\includegraphics[width=1.\columnwidth]{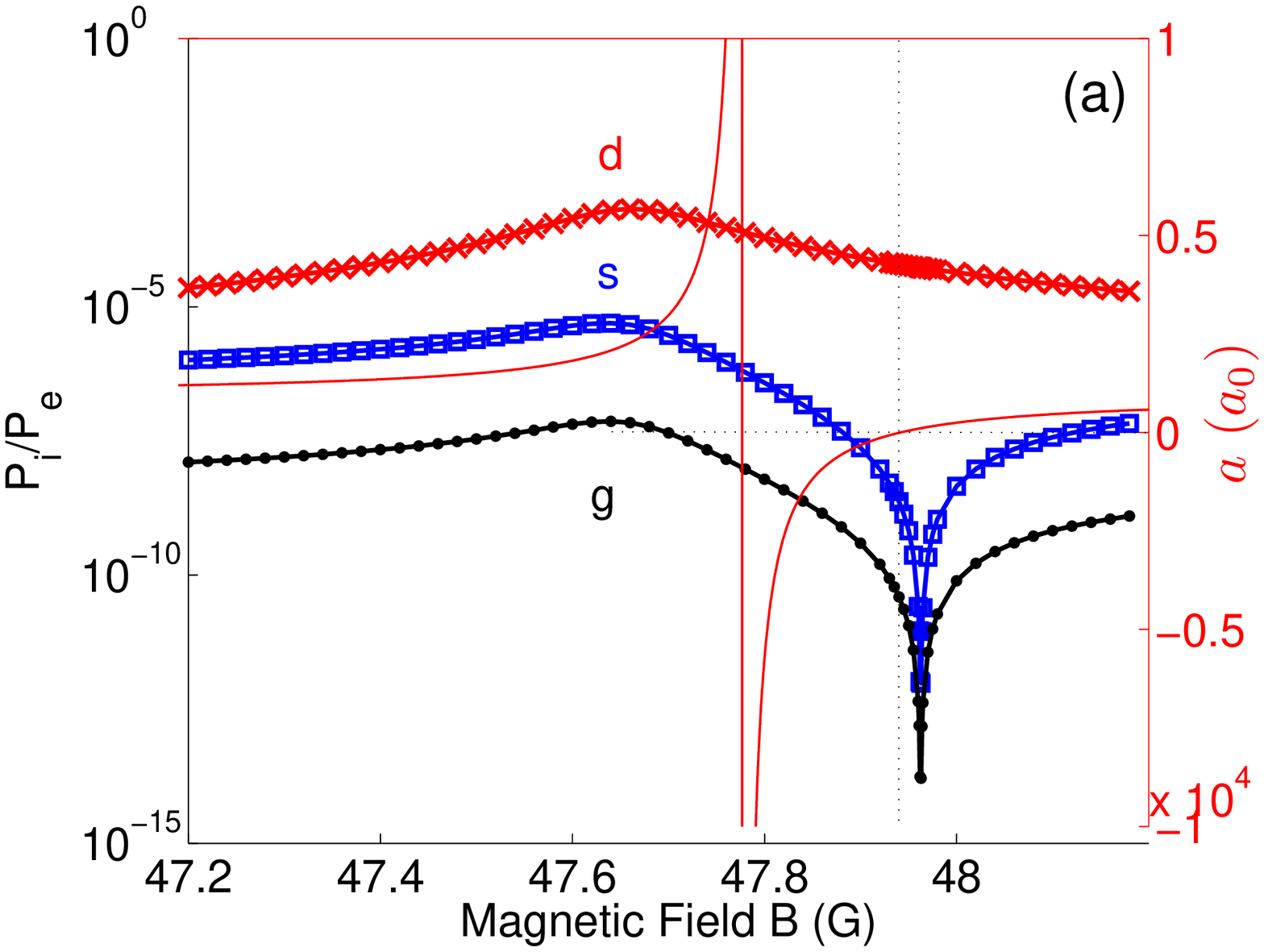}
\includegraphics[width=1.\columnwidth]{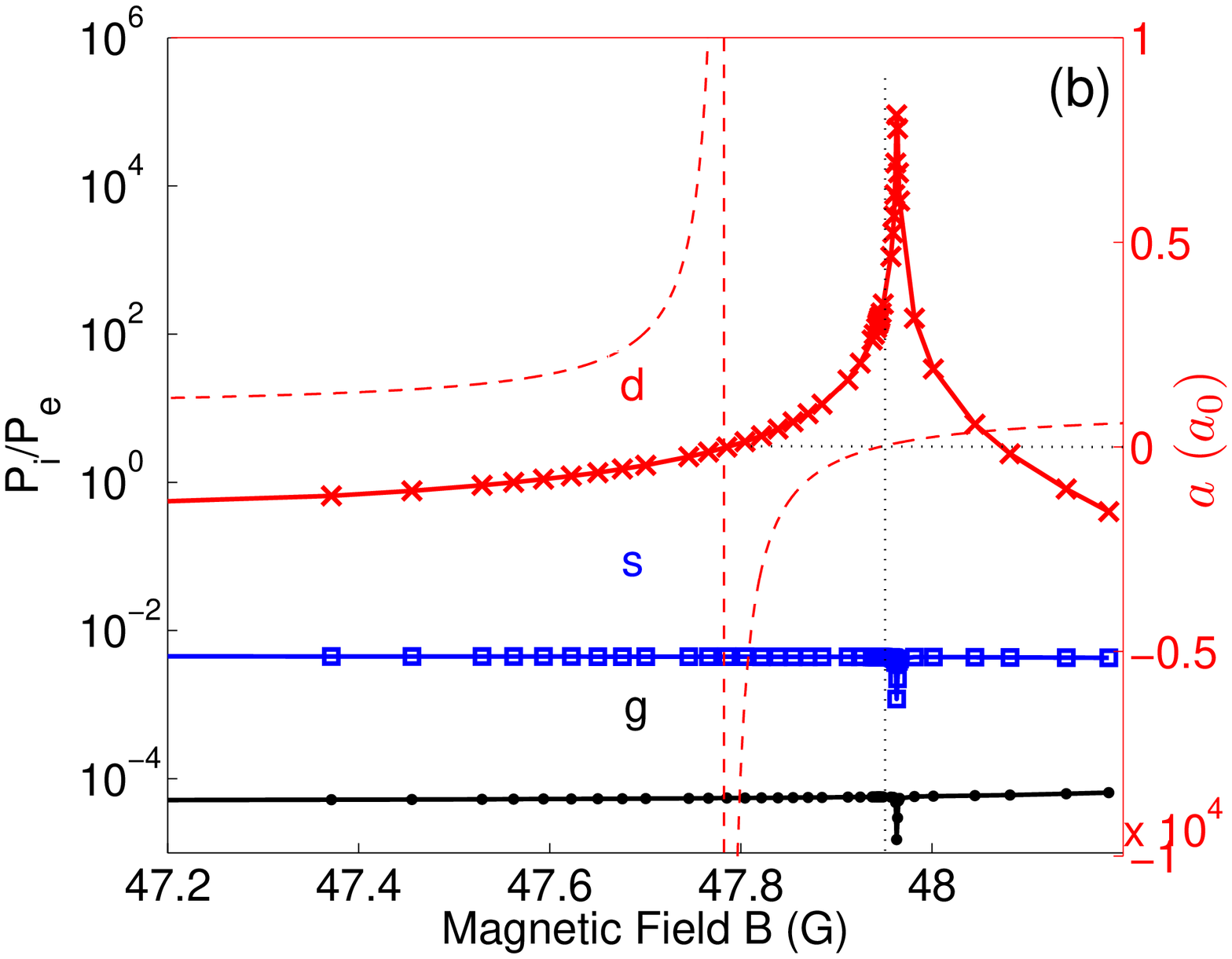}
\caption{(Color online) The relative populations $P_i/P_e$ (13) of the molecular
states $i=s,d$ and $g$  calculated as a function of $B$ near the
d-wave Feshbach resonance in Cs for the pair collisions in free
space (a) and in the harmonic waveguide with
$\omega_{\perp}=59.6$kHz (b). The s-wave scattering length $a(B)$
in free space from Fig.1 is also given.} \label{fig6}
\end{figure}
\begin{figure}[t]
\includegraphics[width=1.\columnwidth]{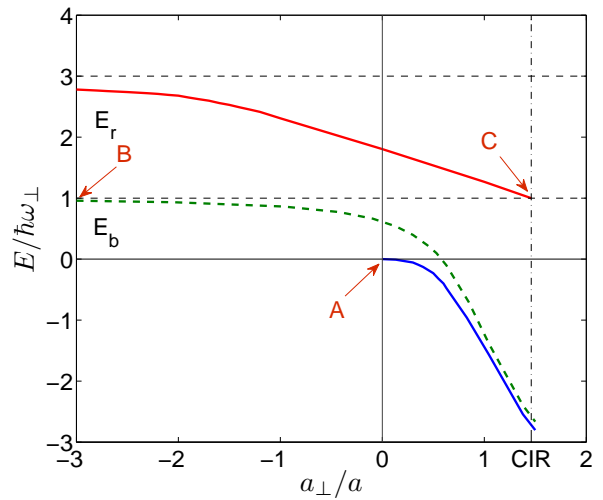}
\caption{(Color online) Schematic picture (presented in \cite{Olshanii2,Melezhik11} for a single-channel potential model of
interatomic interaction) of
the bound state $E_b$ of two atoms as a function of $a_{\perp}/a$
in free space (solid line) and in the harmonic waveguide (dashed
line). The first excited resonance state $E_r$ is also presented.}
\label{fig7}
\end{figure}

To conclude this subsection we also give the expression for the position of the CIR with respect to the resonance at $a=0$ (to the point
$B_i^*$)
\begin{equation}
\Delta_{CIR}^{(i)}= \frac{\Delta_i}{1-C\gamma_i\frac{a_{bg}}{a_{\perp}}} \,\,.
\end{equation}
Since the resonance maximum at $B_i^*$ corresponds to the unit value
$T(B_i^*)\rightarrow 1$ and the minimum of the CIR approaches to
zero $T(B_i^*-\Delta_{CIR}^{(i)})\rightarrow 0$, we can define the ``right"
half width $\frac{1}{2}\Gamma_{CIR}^{(i)}$ of the CIR as the distance of the zero of the
coefficient $T$ to the right situated point where $T=\frac{1}{2}$. Using the formulas for
$\Delta_{CIR}^{(i)}$ and Eqs.(10) for $\Gamma_{i}^*$ we obtain
\begin{equation}
\frac{1}{2}\Gamma_{CIR}^{(i)}=\Delta_{CIR}^{(i)}-\frac{1}{2}\Gamma_i^*
\ _{\overrightarrow{(k_0\rightarrow 0)}}
\Delta_i[\frac{1}{1-C\gamma_i\frac{a_{bg}}{a_{\perp}}}-\frac{a_{\perp}^2
k_0}{2a_{bg}\gamma_i}]\,\,.
\end{equation}
The above expression describes the narrowing of the CIR width
$\Gamma_{CIR}^{(i)}$ with decreasing trap frequency
$\omega_{\perp}$, opposite to $\Gamma_i^*$ (see Fig.4).
This expression might also be of interest to the experimental analysis of the CIRs.

\subsection{Population of molecular states in free space and in the waveguide}

We have also calculated the relative populations $P_i/P_e$ of the
molecular states $|c,i\rangle$ in the process of pair atomic
collisions for harmonic traps of different frequencies
$\omega_{\perp}$ as well as in free space ($\omega_{\perp}=0$).
The populations $P_i$ are defined as
\begin{equation}
P_i=2\pi\int_0^{\infty}\int_0^{\pi}|\psi_{c,i}(r,\theta)|^2
 r^2 d r \sin\theta d \theta\,\,,
\end{equation}
where integration over the infinite region $0 \leq r <\infty$
gives a convergent results due to the decaying tails of the
molecular bound-state wave-functions $\psi_{c,i}({\bf
r})\rightarrow 0$ at distances of the order $\sim\overline{a}$ in
the closed channels $|c,i\rangle$. The population $P_e$ is defined
in a region near the origin $r\rightarrow 0$ of the molecular
dimension $\sim\overline{a}$  in the entrance channel by using
Eq.(13) where the upper limit of the integration over $r$ is
$\overline{a}$. The relative populations $P_i/P_e$ of the
molecular states for ultracold atomic collisions in free space and
in the harmonic waveguide with $\omega_{\perp}=59.6$kHz are shown
in Fig.8 in the region of B near the d-wave Feshbach resonance. In
free space the pair collision leads to a relatively low population
of the d-wave molecular state $P_2/P_e\sim 10^{-4}-10^{-5}$. The
populations of non-resonant s- and g-wave states are $P_1 \ll P_2$
and $P_3 \ll P_2$ essentially suppressed here. The dependence of
the d-wave molecular state population on $B$ repeats the
dependence on $B$ of the population of the region
$\sim\overline{a}$ in the entrance channel $P_2(B)/P_e(B)\sim
const$. However, in the confined geometry of the waveguide
($\omega_{\perp}\neq 0$) the pair collision leads to a resonant
enhancement of the relative population $P_2(B)/P_e(B)$ of the
d-wave molecular state near the point $B_2^*=47.944G$ where the
free space scattering length is  zero $a(B_{2}^*)=0$. In the
waveguide the point of appearance of the bound state is shifted
from the position defined by $1/|a|\rightarrow 0$ for free space
scattering (see point A in the illustrative scheme of the bound
and resonant states of an atomic pair given in Fig.9) to the point
$1/|a|\rightarrow \infty$ (the point B in Fig.9). This is why we
observe in Fig.8(b) the strong resonant enhancement of the
population $P_2(B_{2}^*)$ at the magnetic field $B_{2}^*=47.944$G.
The populations $P_1/P_e$ and $P_3/P_e$ of other molecular states
also show some enhancement with respect to the entrance channel
due to the coupling of the states with the ``resonance" channel
$|c,2\rangle$, which, however, are a few orders of magnitude less
than the enhancement of the population $P_2/P_e$. We do not
observe a resonant behavior of $P_2(B)$ at the point $B=47.57$G of
the CIR (see Fig.4 and the point C in Fig.9). We suspect that this
is the case due to the rather weak coupling in our model potential
$\hat{V}$(2) between the entrance channel $|e\rangle$ and the
resonant state in the closed channel $|c,2\rangle$ for stimulating
considerable transition to the molecular state $|c,2\rangle$ in
the closed excited channel of the waveguide.

In Figs.10  and 11 we present the results of our computation of
the molecular state populations near the s- and g-wave Feshbach
resonances -11.1G and 53.449G, respectively. We observe the
qualitatively analogous effect of resonant enhancement of the
relative populations $P_1(B)/P_e(B)$ and $P_3(B)/P_e(B)$ at the
points $B_{1}^*=18.1$G and $B_{3}^*=53.457$G in the harmonic
waveguide, which correspond to the positions of the zero of the
s-wave scattering length in the vicinity of s- and g-wave Feshbach
resonances, respectively. Here we notice $P_1(B)/P_e(B)\simeq const$, i.e.
the same dependence on $B$ of the s-wave molecular state population as
the population of the region $\leq\overline{a}$ in the entrance
channel $P_1(B)/P_e(B)\simeq const$ (see Fig.10(a)). In Fig.11(a) we
observe the resonant enhancement of the relative population
$P_3(B)/P_e(B)$ of the g-wave molecular state near the Feshbach
resonance in free space at $B_{0,3}=53.449G$ where $1/|a|\rightarrow
0$. A similar resonance enhancement near the
point $1/|a|\rightarrow 0$ of the appearance of the near threshold
resonance or weakly-bound state in free space (see point A in
Fig.9) is observable via the values $P_1(B)$,$P_2(B)$ and $P_e(B)$ but,
because $P_1(B)/P_e(B)\simeq const$ and $P_2(B)/P_e(B)\sim const$,
we do not observe this effect in the relative populations
$P_1(B)/P_e(B)$ and $P_2(B)/P_e(B)$ in Figs.10(a) and 8(a).

\begin{figure}[hbt]
\includegraphics[width=1.\columnwidth]{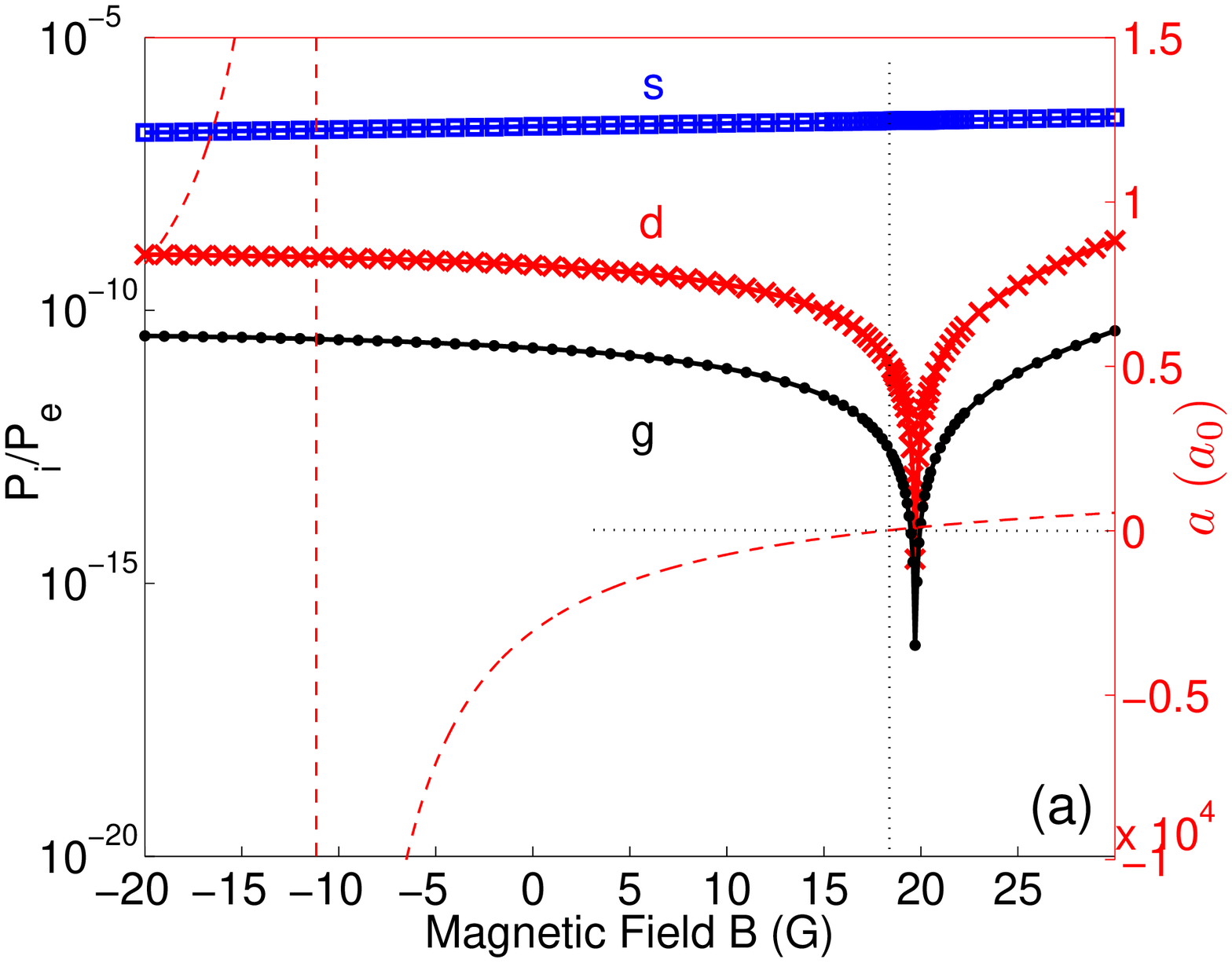}
\includegraphics[width=1.\columnwidth]{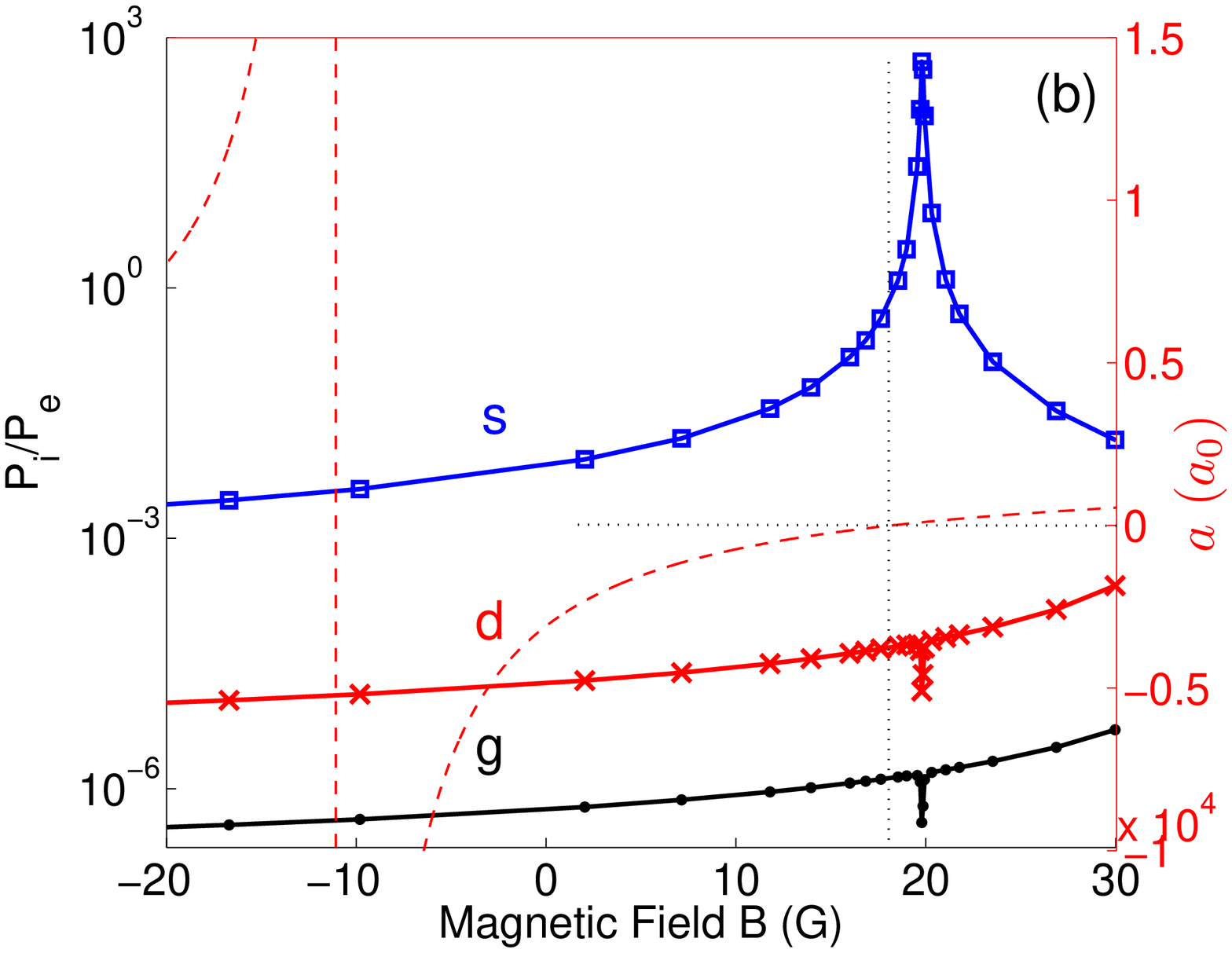}
\caption{(Color online) The relative populations $P_i/P_e$ (13) of the molecular
states $i=s,d$ and $g$  calculated as functions of $B$ near the
s-wave Feshbach resonance for Cs for the pair collisions in
free space (a) and in the harmonic waveguide with
$\omega_{\perp}=59.6$kHz (b). The s-wave scattering length $a(B)$
from Fig.1 is also provided.} \label{fig8}
\end{figure}

\begin{figure}[hbt]
\includegraphics[width=1.\columnwidth]{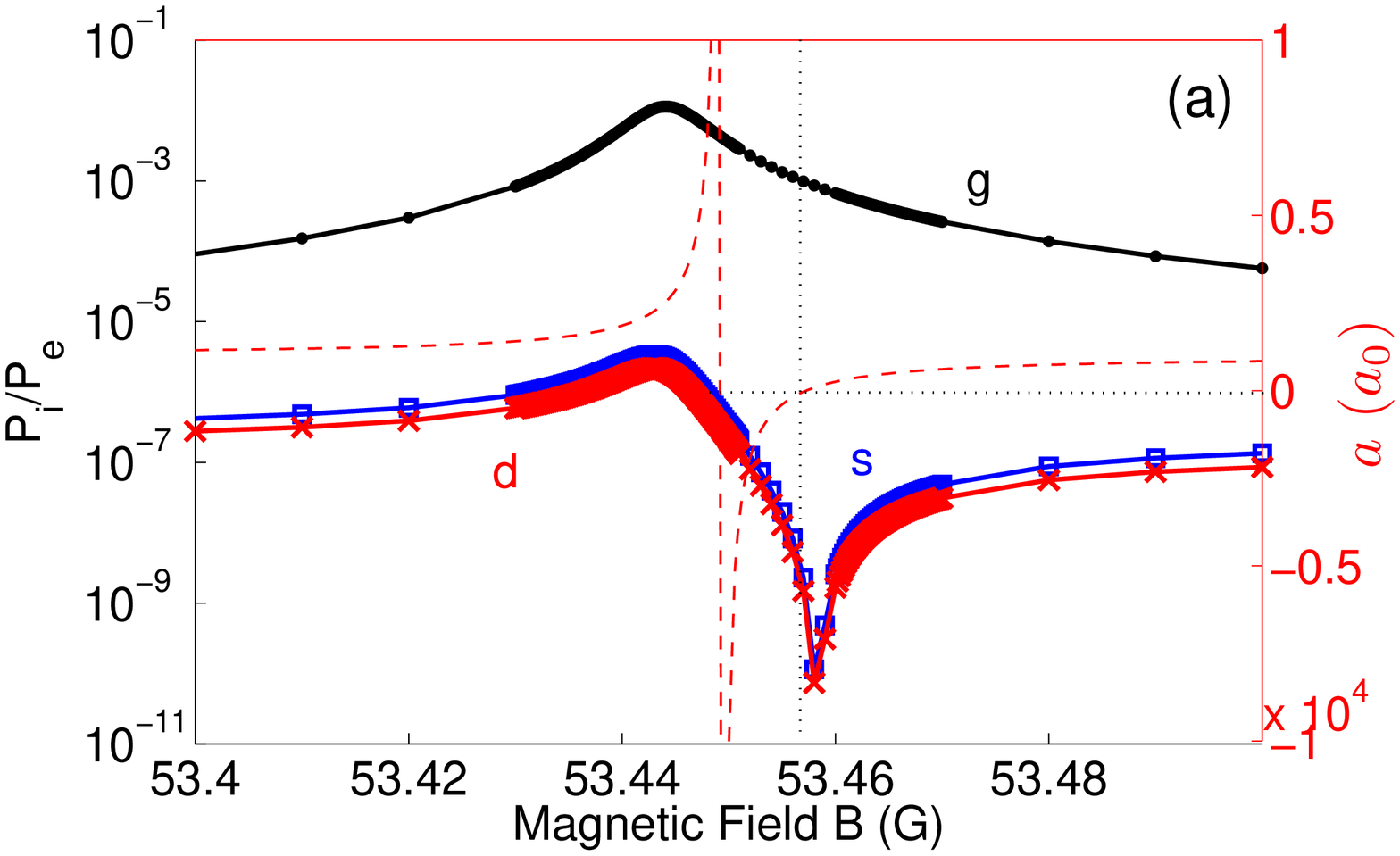}
\includegraphics[width=1.\columnwidth]{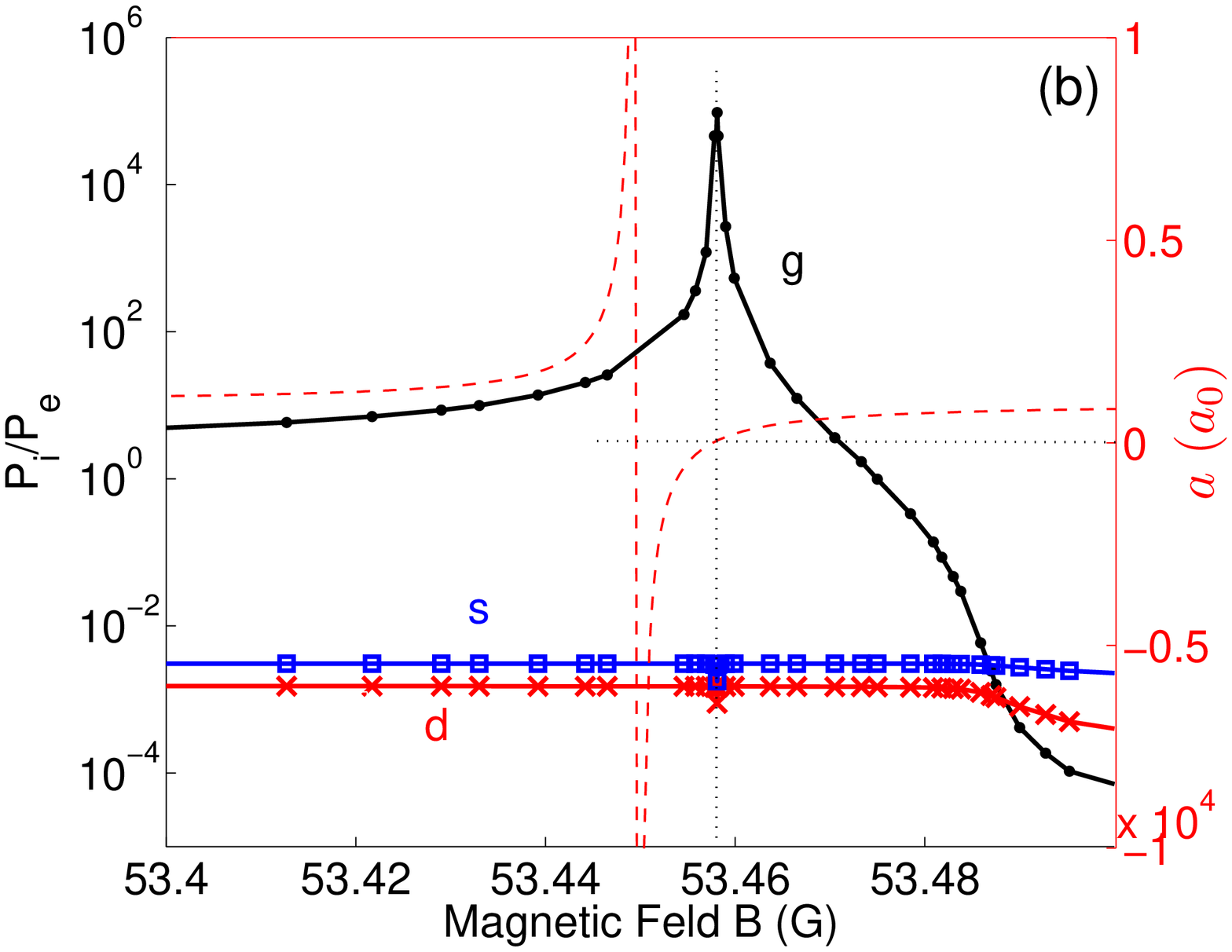}
\caption{(Color online) The relative populations $P_i/P_e$ (13) of the molecular
states $i=s,d$ and $g$  calculated as functions of $B$ near the
g-wave Feshbach resonance in Cs for the pair collisions in
free space (a) and in the harmonic waveguide with
$\omega_{\perp}=59.6$kHz (b). The s-wave scattering length $a(B)$
from Fig.1 is also presented.} \label{fig9}
\end{figure}

\section{CONCLUSION}
We have developed a theoretical model for a
quantitative analysis of the Feshbach resonance shift and width
induced by an atomic waveguide. It is based on our multichannel
approach for confinement-induced resonances and atomic
transitions in the waveguides in the multi-mode regime
\cite{Saeidian}. In this scheme the single-channel
(scalar) interatomic interaction is replaced by a four-channel (tensorial)
potential modeling resonances of different structure according
to the two-channel parametrization of A.D.Lange et. al.
\cite{Lange}. The experimentally known parameters of
the Feshbach resonance in the absence of the waveguide are used
as an input in our approach.
We have calculated the shifts and widths of s-, d- and g-wave
magnetic Feshbach resonances of Cs atoms emerging in harmonic
waveguides as CIRs and resonant enhancement of the transmission at zeros of the free space scattering length.

In particular we find that the relationship $a=a_{\perp}/C$ for the position of
the CIR in a harmonic waveguide is fulfilled with high accuracy
for the Feshbach resonances of different tensorial structure which
holds in spite of the fact that this property was originally obtained in the framework of a s-wave
single-channel pseudopotential approach \cite{Olshanii1}. Note,
that this property was experimentally confirmed for d-wave Feshbach
resonances in a gas of Cs atoms \cite{Haller2010}. The maximum
of the transmission, corresponding to the zero of the scattering
amplitude, is shown to be independent of the trap field strength
and then again corresponds to the zero $B_{i}^*$ of the s-wave scattering length $a(B)$
in free space. In a nutshell, the Feshbach resonance in
free space develops in the harmonic waveguide into a minimum of $T$ (position
of CIR), defined by the formulas $a=a_{\perp}/C$, and a maximum,
coinciding with the position of zero of the s-wave scattering
length a. The 'distance' between these extrema is equal to $\Delta_{CIR}^{(i)} =\Delta_i/(1-C\gamma_i
a_{bg}/a_{\perp})$.

We have derived expressions for the widths $ \Gamma_{i}^*,\frac{1}{2}\Gamma_{CIR}^{(i)}$ of the resonant enhancement
of $T$ at $B_i^*$ and the ``right" side half-width of the minimum of the
$T$-coefficient i.e. at the position of CIR and confirmed its
validity by numerical results for $k_0\rightarrow 0$. By
measuring the width $ \Gamma_{i}^*$ one can, in principle, extract from these expressions
the longitudinal collision energy and estimate the ``longitudinal" temperature of
the atomic cloud in the trap. In other words the
width of the atomic loss resonance observed in the experiment
\cite{Haller2010} at the point of CIR might contain important
information about the longitudinal atomic momentum $k_0$ and the
temperature of the gas. It also shows that one can control the
width $\Gamma_{i}^*$ of the resonance at $a(B_i^*)=0$ by varying the trap
frequency $\omega_{\perp}$. An increase of $\omega_{\perp}$ leads
to a narrowing of the resonance, an effect which could potentially
be used experimentally.

Finally, the molecule formation rates in a waveguide show an enhancement
for the case of a corresponding zero of the s-wave scattering
length $a(B_{i}^*)=0$. We have shown that the positions of these
resonances are stable w.r.t. the variation of the confining frequency
$\omega_{\perp}$ of the waveguide.

Our model adds to the possible studies of
scattering processes of ultracold atomic gases in waveguides
beyond the framework of s-wave resonant scattering. Our model
might be extended to the cases of fermions or distinguishable atom
scattering, including transverse excitation/diexcitation processes \cite{Saeidian}.
It permits also for the investigation of other trap geometries
\cite{MelHu} and more realistic interatomic interactions.

\section{ACKNOWLEDGEMENTS}
We thank E. Haller and H.-C. N\"agerl for fruitful discussions. Authors
V.S.M. and P.S. acknowledge financial support by the Deutsche
Forschungsgemeinschaft and the Heisenberg-Landau Program. V.S.M.
thanks the Zentrum f\"ur Optische Quantentechnologien of the
University of Hamburg and Sh.S. thanks the Bogoliubov Laboratory of
Theoretical Physics of JINR at Dubna for their warm hospitality.

\end{document}